\documentstyle[a4wide,12pt,epsfig,fancyheadings]{article}
\newcommand{\be}{\begin{equation}}
\newcommand{\ee}{\end{equation}}
\title{Bloch-Nordsieck Summation and Partonic Distributions in Impact Parameter
Space}
\author{A. Corsetti\\
INFN, Physics Department, University of Rome La Sapienza, Rome, Italy\\
A. Grau\\Departamento de F\'\i sica Te\'orica y del Cosmos,
Universidad de Granada, Spain\\
 G. Pancheri \\
INFN - Laboratori Nazionali di Frascati , I00044 Frascati\\
and\\
 Y.N. Srivastava\\
Physics Department and INFN, University of Perugia, Perugia, Italy}
\date{ }
\begin{document}
\maketitle
\thispagestyle{fancy}
\begin{abstract}

A model for the parton distributions of hadrons in impact parameter space 
  has been 
constructed using soft gluon summation. This model 
incorporates the salient features of
distributions obtained from the intrinsic transverse momentum
behaviour of hadrons. Under the assumption that the intrinsic behaviour is
dominated by soft gluon emission  stimulated by the scattering process,
 the b-spectrum becomes softer and softer as the
 scattering energy increases. In minijet models for the inclusive 
  cross-sections, 
 this will counter the increase from $\sigma_{jet}$ .
\end{abstract}

The impact of parton scattering  on the rise with energy 
of inclusive cross-sections 
was suggested  by Cline and Halzen \cite{CLINE}, after such rise was 
first observed in proton-proton
collisions at ISR .  Minijet 
\cite{CR,GAISSER,PS} and eikonal minijet models \cite{DURAND,BLOCK,TRELEANI,
CAPELLA} 
were subsequently developed to include 
 an increasing number of partonic collisions in QCD
resulting from the rapid rise in gluon densities.  
Recent measurements of photo- and hadro-production total
 cross-sections
\cite{HERA,TEVA} in 
energy regions where QCD processes dominate, confirmed the trend and were
confronted  with
theoretical predictions obtaining varying degrees of success \cite{GODBOLE}. 
Typically, the eikonal
(unitary) models for total hadron-hadron cross-sections are written
as
\be
\label{eiksigma}
\sigma_{total}=2\int d^2{\vec b} \big[ 1-e^{-n(b,s)/2}\big]
\ee
where the average number of collisions at impact parameter b is given by
\be
\label{nb}
 n(b,s)=A(b)\sigma (s)
\ee
In mini-jet models,  $\sigma (s)=\sigma_{non-pert}(s)+\sigma_{jet}(s)$, with
$\sigma_{jet}$ to be calculated  from perturbative QCD.
 A key ingredient of all models with a QCD component
is the overlap function  $A(b)$, which describes 
matter  distribution in  impact parameter space. In some models
\cite{DURAND}, it
 has been assumed to be
the Fourier transform of the product of  hadronic form factors of the colliding 
particles. 
In other models \cite{PYTHIA}, a gaussian shape has been assumed, 
thus relating A(b) to the 
intrinsic transverse momentum distribution of  partons in the colliding
hadrons. In either case, 
detailed information on $A(b)$ relies on parameters
to be determined  case by case. Moreover, while 
direct measurements of the EM form factors are  available for nucleons 
and pseudoscalar mesons, experimental information regarding photons or
other 
hadrons such as vector mesons is lacking. The same observation applies
to the intrinsic-$p_t$ interpretation for the spatial distribution of partons 
in vector mesons or photons. There is clearly need for a theoretical 
description of
these distributions which would allow for  a calculation of the mini-jet total 
cross-section (in the eikonal approximation) for photo-production and   
gamma-gamma collisions at future colliders.

\thispagestyle{plain}
The aim of this paper is  to provide a model for $A(b)$ which
could be applied to various cases of interest.  We shall use  
Bloch-Nordsieck techniques to sum soft gluon transverse momentum distributions
 to all orders and compare 
our results with both the intrinsic transverse momentum approach as well as
the form factor approach. In what follows, we shall first illustrate
the Bloch-Nordsieck result and show that  it gives 
a gaussian fall-off with an intrinsic transverse size consistent with
Montecarlo models \cite{PYTHIA}.
We then calculate the 
relevant distributions and discuss their phenomenological
application.

We propose that in hadron-hadron collisions,
 the b-distribution of 
partons in the colliding hadrons is the Fourier transform of the 
transverse momentum distribution resulting from
soft gluon radiation emitted by quarks as the hadron 
breaks up because
of the collision. This distribution is obtained by summing soft 
gluons to all orders,
with a technique amply discussed in the literature \cite{EPT,GPSS}. 
The resulting
expression is \cite{DDT,PP}
\thispagestyle{plain}
\begin{equation}
\label{BNPT}
{\cal F_{BN}}(K_\perp)={{1}\over{2 \pi}}\int b db J_0(b K_\perp) e^{-h(b;M,
\Lambda)}
\end{equation}
with 
\begin{equation}
\label{hdb}
h(b;M,\Lambda)={{2 c_F}\over{\pi}}\int_0^M {{dk_\perp}\over{k_\perp}}\alpha_s({{k^2_\perp}
\over{\Lambda^2}})\ln{{M+\sqrt{M^2-k_\perp^2}}\over{M-\sqrt{M^2-k_\perp^2}}}
[1-J_0(k_\perp b)]
\end{equation}
 The choice
of the hadronic scale M, which accounts for the maximum energy allowed in
a single ($k^2=0$) gluon emission, plays a crucial 
role, just as it did for the Drell-Yan  process, where the above
 expression has been successfully 
\cite{GRECO,HALZEN,NAKW,ALTARELLI} used to describe the transverse momentum  
distribution of the
time-like virtual photon  or W-boson. In these cases\cite{GRECO,NAKW},
the scale M was found to be energy
dependent and to vary between $\sqrt{Q^2}/4$ and $\sqrt{Q^2}/2$. We shall 
return to this determination later.

\thispagestyle{plain}
 The definition given in eqs.(\ref{eiksigma},\ref{nb}), requires for its 
consistency a  normalized  b-distribution, i.e.  
\be
\label{norm} 
\int d^2 {\vec b} A(b) = 1.
\ee 
Our proposed Bloch-Nordsieck expression for the overlap function
 $A(b)$, satisfying the above normalization, reads
\be
\label{ourAB}
A_{BN}= {{e^{-h(b;M,\Lambda)}}\over
{2\pi \int bdb 
 e^{-h(b;M,\Lambda)}}}
\end{equation}
An inspection of eq.(\ref{hdb}), immediately 
poses the problem of extending the known asymptotic freedom expression 
for $\alpha_s$ to the very small $k_\perp$ region. 
To avoid the small $k_\perp$ divergence in eq.(\ref{hdb}), 
it has been customary to   introduce a lower cut-off in $k_\perp$ 
and freeze $\alpha_s$ at $k_\perp=0$, i.e. to put
\begin{equation}
\label{altarelli}
\alpha_s(k_\perp^2)={{12 \pi}\over{33-2 N_f}} {{1}\over{\ln[(k_\perp^2+a^2 
\Lambda^2)/\Lambda^2)]}}
\end{equation}
with $a=2$ in ref. \cite{ALTARELLI}. For applications where the
scale $M$ is large  (e.g., W-transverse momentum distribution calculations)
eq.(\ref{hdb}) is dominated by the (asymptotic) logarithmic behaviour and 
the small $k_\perp$-limit, albeit theoretically crucial, is not very relevant 
phenomenologically. However, this is not case in the present context,
where we are dealing with soft gluon emission in low-$p_t$ physics 
(responsible for large cross-sections). The typical scale 
of such peripheral interactions,  is that of the hadronic masses, i.e.
we expect $M\sim {\cal O}(1\div2 \ GeV)$ and the small $k_\perp$ limit
plays a basic role. This can be  appreciated  on a qualitative basis, by 
considering the limit  $ b M<<1$ of eq. (\ref{hdb}).
  In this region, we can approximate 
$1-J_0(kb)\approx b^2k^2/4$, to obtain
\begin{equation}
\label{intkdk}
h(b;M,\Lambda)\approx b^2\ A
\ee
with 
\be
\label{A}
 A= {{c_F}\over{4\pi}}
\int  dk^2 \alpha_s({{k^2}\over{\Lambda^2}}) \ln
{{4 M^2}\over{k^2}}
\end{equation}
We obtain a function $h(b;M,\Lambda)$ 
with  a gaussian fall-off as in models where 
  $A(b)$ is the Fourier transform of an intrinsic transverse momentum 
  distribution of partons, i.e.  
  \newline $\exp(-k_\perp^2/4 A^2)$. Note that the relevance of an integral
 similar
to the one in eq.(\ref{A}) has been recently discussed in connection to
 hadronic event shapes \cite{DW}.

\thispagestyle{plain}
 Our choice for the infrared behaviour of $\alpha_s$ for a
  quantitative description of the  distribution in eq. (\ref{hdb}),
  does not follow eq. (\ref{altarelli}), but  is inspired by the
   Richardson potential
  for quarkonium bound states \cite{RICHARDSON}. In 
 a number of related
applications \cite{INTRDY,ourWBOSON}, we have proposed to calculate 
the above integral using the  following expression for $\alpha_s$ :
\begin{equation}
\label{alphaRich}
\alpha_s(k_\perp)={{12 \pi }\over{(33-2N_f)}}{{p}\over{\ln[1+p({{k_\perp}
\over{\Lambda}})^{2p}]}}
\end{equation}
which coincides with the usual one-loop expression for large (relative
to $\Lambda$) values of $k_\perp$, while going to a singular
 limit for small $k_\perp$. For the special case $p=1$ such an
$\alpha_s$ coincides with one used in the 
 Richardson potential \cite{RICHARDSON},
and which  incorporates - in a compact expression - 
 the high-momentum limit demanded by asymptotic freedom as well as 
 linear quark confinement in the static limit. We have
generalized Richardson's ansatz to values of $p\le 1$. For $1/2< p\le 1$,
this corresponds to a confining potential rising less than linearly
with the interquark distance $r $. 
The range $p\neq 1$ has an important advantage, 
i.e., it allows the integration in eq.(\ref{hdb}) to converge for all 
values of $k_\perp$. A qualitative argument to justify the use 
of such (less singular) values for the parameter p  follows.

\thispagestyle{plain}
Assume a confining potential (in momentum space) given by ``one-loop gluon'' 
exchange term 
\begin{equation}
{\tilde V}(Q) = K ({{\alpha_s(Q^2)}\over{Q^2}}),
\end{equation}
where K is a constant calculable from the asymptotic form of $\alpha_s$. Let
us choose the simple form 
\begin{equation}
\alpha_s(Q^2) = {{1Ê}\over{b\ \ln [1 + (Q^2/\Lambda^2)^{p}]}},
\end{equation}
so that ${\tilde V}(Q)$ for small Q goes as 
\begin{equation}
\label{potential}
{\tilde V}(Q) \rightarrow  Q^{-2(1 +p)}.  
\end{equation}
Eq.(\ref{potential}) implies, for the potential, in coordinate space
\begin{equation}
V(r) =\ \int{{d^3Q}\over{(2\pi)^3}} e^{i {\bf Q.r}} {\tilde V}(Q),  
\end{equation}
as
\begin{equation}
V(r) \rightarrow (1/r)^3 \cdot r^{(2 + 2p)} \sim  C\ r^{(2p - 1)},
\end{equation} 
for large r (C is another constant). A simple check is that for $p$ 
equal to zero, the usual Coulomb potential is regained. Notice that for
a potenial rising with r, one needs $p > 1/2$. A determination of $p$ 
can be made following an argument due to Polyakov \cite{POL}. We
minimize the potential, consistent with the constraint that the
potential energy (or mass, M) increases as $J^{(1/2)}$ for large J. ( A 
linear Regge trajectory means that the angular momentum $J\ \sim M^2$). 
The ``confinement'' energy plus the rotational energy ( all for large r and 
large J) reads
\begin{equation}
V(r,J)\ \sim ({{J(J+1)}\over{r^2}})\ +\ C r^{(2p - 1)}. 
\end{equation}
Since the first term goes down with r and the second grows with r (for
confinement, $p\ >\ (1/2)$ is necessary), a minimum is
guaranteed (for a fixed J). So we minimize the above potential, 
and find the position $r\ =\ r_o(J)$ where the minimum is, as a function 
of J, 
\begin{equation}
r_o(J) \sim J^{2/(2p +1)}, \hbox{ for large}\ J.  
\end{equation}
\thispagestyle{plain}
Inserting this value of $r_o$, we find the minimum effective potential
(or rest mass) as a function of J
\begin{equation}
U_{min}(J,r_o) \sim J^{({{4p -2}\over{2p +1}})}.
\end{equation}
A mass spectrum with linearly rising Regge trajectories, for which 
$M^2 \sim J$, implies  the right side to grow like $J^{(1/2)}$, which gives
\begin{equation}
p \ =\ (5/6).  
\end{equation}

\thispagestyle{plain}
For the motivations given in \cite{INTRDY} and repeated above,
the value $p=5/6$ was chosen  in previous calculations of  
the transverse momentum distribution of Drell Yan pairs \cite{INTRDY,NAKW}. 
Compared to  eq.(\ref{altarelli}), we consider our present choice
 (i.e., eq.(\ref{alphaRich})) physically more transparent as 
 it  leads to a direct quantitative estimate of soft-gluon
generated intrinsic transverse momentum of partons in hadrons. 
 \begin{figure}[h]
\begin{center}
\leavevmode
\mbox{\epsfig{file=partonspacefig1.ps,width=0.7\textwidth,bbllx=30pt,bblly=70pt,bburx=570pt,bbury=750pt,angle=90}}
\end{center}
\caption{Intrinsic transverse momentum of partons 
in a hadron for different hadron scales M, in units of $\Lambda$, for p=5/6.}
\label{intrinsic}
\end{figure}
With our ansatz, the intrinsic transverse momentum becomes a calculable 
quantity rather than being an assumed one (see related discussion in 
Nakamura et al. \cite{INTRDY}). It can be obtained from the decomposition
\begin{equation}
\label{hdecomposed}
h(b;M,\Lambda)=h_{intrinsic}+\Delta(b;M,\Lambda)
\end{equation}
with
\begin{equation}
h_{intrinsic}={{32}\over{33-2N_f}}\int_0^{\Lambda} 
{{d k}\over{k}}{{p}\over{\ln(1+p({{k^2}\over{\Lambda^2}})^{p})}}
\ln{{M+\sqrt{M^2-k^2}}\over{M-\sqrt{M^2-k^2}}}(1-J_0(bk)),
\end{equation}
where  the contribution to the integral from the region 
$k_\perp \le \Lambda$ determines the form and type of the ``intrinsic 
transverse momentum'' behaviour, as we discussed above. 
\thispagestyle{plain}
In this region, we now have 
\begin{equation}
h_{intrinsic}\approx {{b^2}\over{4}}<k_t^2>_{int}
\end{equation}
with
\begin{equation}
<k_t^2>_{int}={{32}\over{33-2N_f}}{{1}\over{(1-p)}}
\left({{1}\over{2(1-p)}}+\ln{{2M}\over{\Lambda}}\right)\Lambda^2,
\end{equation}
By comparison with  eq.(\ref{intkdk}), we see that
$A=<k_t^2>_{int}/4$, 
which corresponds to  an intrinsic transverse  momentum of a few hundred MeV for
$\Lambda$ in the 100 MeV range and $M\le 1$ GeV. For $N_f=3$, we show in
Fig.1 the variation of $\sqrt{<k_t^2>_{int}}$
as a function of $M$ for a range of values of $\Lambda$.

\thispagestyle{plain}

The ``intrinsic'' behaviour just discussed appears in the very small 
$K_\perp$ region, i.e. $K_\perp \le \Lambda$. In general, 
 the full expression eq.(\ref{hdecomposed}) 
should be used.

Having set up our formalism, we shall now examine  its implications. 
The distribution $A(b)$ depends upon the hadronic
scale M in the function $h(b)$. This scale
   depends upon the energy of
the specific subprocess
and, through this, upon  the hadron scattering energy. In the
calculation of the transverse momentum distribution of a lepton
pair produced in  quark-antiquark
annihilation \cite{GRECO},
the function $e^{-h(b)}$ was the Fourier-transform of such distribution
and 
the scale $M$ was obtained as
  the maximum transverse momentum allowed by
kinematics to a single gluon emitted by the initial $q{\bar q}$ pair of
 c.m.energy 
$\sqrt{{\hat s}}$ in the process
\be
\label{processDY}
q {\bar q} \rightarrow g + \gamma(Q^2)
\ee
Following ref.\cite{GRECO}, for a lepton pair
of mass squared $Q^2$ and rapidity $y$,
this quantity is given by
\be
\label{qmax}
q_{max}({\hat s},y)={{ \sqrt{{\hat s}} }\over{2}}
 (1-{{Q^2}\over{{\hat s}}})
{{1}\over{\sqrt{1+zsinh^2y}}}
\ee
with $z=Q^2/{\hat s}$. This would be the upper limit of integration in the 
function
$h(b)$ for each particular subenergy ${\hat s}$ of the quark-antiquark pair.
The transverse momentum distribution can then be factorized from the
 overall
hadronic cross-section, by substituting M 
 in $h(b)$ with the average of $q_{max}$
over all quark energies and configurations, with a weight
$1/{\hat s}$, proportional to the cross-section. For a fixed value
of the Feynman variable $x_F$, one obtains
\be
\label{M}
M(s,x_F)={{\int {{dx_1}\over{x_1}}\int{{dx_2}\over{x_2}}\delta(x_1-x_2-x_F)
q_{max} f(x_1)f(x_2)}\over{ \int {{dx_1}\over{x_1}} \int {{dx_2}\over{x_2}}
\delta(x_1-x_2-x_F)f(x_1)f(x_2)}}
\ee
for given quark densities $f(x)$.
To understand the energy dependence of this scale, one can use
a simple toy model, in which one
 approximates the valence quark densities 
 with $1/\sqrt{x}$ and 
obtain
from, eq.(\ref{M}) at $x_F=0$,
\be
\label{mdy}
M(s,0)\approx {{2Q}\over{3}} 
\left(1-3{{Q}\over{2\sqrt{s}}}\right)\ \ \ \ Q<<\sqrt{s}
\ee
which shows that M increases with energy towards the asymptotic value $2Q/3$.
This tendency was confirmed by numerical estimates with
realistic quark densities \cite{ourWBOSON}. 
An  asymptotic increase of M with energy is also obtained for the case
of interest here, i.e. parton-parton scattering contribution to the
total cross-section.
In the Drell-Yan case, one needed $h(b)$ to calculate
 the transverse momentum distribution
of the  lepton pair, here we use it to evaluate the average number of
partons in the overlap region  of two colliding hadrons. In this case 
$e^{-h(b)}$ is the {\cal F}-transform of the transverse momentum distribution
induced by initial state radiation  in the process 
\be
\label{process}
q{\bar q} \rightarrow  \ jet \ jet  + X
\ee
where X can also  include the quark-antiquark pair which continues
undetected after  emission of a gluon pair   which stimulated 
the initial state bremsstrahlung. The jet pair   in process 
(\ref{process}) is the one produced through gluon-gluon or other 
parton-parton scattering with total jet-cross-section $\sigma_{jet}$.
  We work in a 
no-recoil approximation, where the transverse momentum of
 the jet pair is balanced by the emitted soft gluons.
 Then the maximum transverse momentum allowed to a single gluon
is still given by an expression similar to eq.(\ref{qmax}), i.e.
\be
\label{qmaxjets}
q_{max}({\hat s})={{ \sqrt{{\hat s}} }\over{2}}
 (1-{{{\hat s_{jet}}}\over{{\hat s}}})
\ee
except that now
$\sqrt{Q^2}$ has been replaced by the jet-jet invariant mass 
$\sqrt{{\hat s_{jet}}}$, over which
one needs to perform  further integrations. In other words, this 
kinematic limit is the same as the one
obtained for the  gluon accompanying the 
Drell-Yan process mentioned above : 
there  
the quark-antiquark pair  annihilates into a dilepton pair and a soft gluon,
here 
 it may continue on its way- undetected- after 
 semi-hard emission of a parton pair which stimulated
the soft gluon initial state 
bremsstrahlung. 
\thispagestyle{plain}
An improved eq.(\ref{nb}) now reads
\be
\label{nbimp}
n(b,s)=n_{soft}(b,s)+\sum_{i,j,}\int{{dx_1}\over{x_1}}
\int{{dx_2}\over{x_2}} f_i(x_1)f_j(x_2)\int dz \int dp_t^2
A_{BN}(b,q_{max}){{d\sigma}\over{dp_t^2 dz}}
\ee
where $f_i$ are the quark densities
 in the colliding hadrons, 
  $z={\hat s_{jet}}/(sx_1x_2)$,  and ${{d\sigma}\over{dp_t^2 dz}}$ is the
  differential cross-section for process (\ref{process})
   for a given $p_t$ of the
  produced jets. In absence of a precise prescription of
  how to deal with the non-perturbative contribution to
  $n(b,s)$, it is customary in these eikonal models
  to introduce a lower cut-off in the jet transverse momentum, usually
  indicated as $p_{tmin}$ to separate the mini-jet contribution from the
  soft part of the cross-section. 
\thispagestyle{plain}
  
Unlike the usual expressions for $n(b,s)$, eq.(\ref{nbimp}) does not
exhibit factorization between the longitudinal and transverse
degrees of freedom since the distribution $A_{BN}$ depends upon the
quark subenergies. Factorization can be obtained however, through an
averaging process similar to the one described above for
the transverse momentum distribution of Drell-Yan pairs :
one     can factorize 
  the b-distribution in eq.(\ref{nbimp}), 
  by evaluating $A_{BN}$ with $q_{max}$ at its mean value, i.e. write
  \be
\label{nbimpmean}
n(b,s)=n_{soft}(b,s)+A_{BN}(b,<q_{max}(s)>)\sigma_{jet}
\ee
with
\be
\sigma_{jet}=\sum_{i,j,}\int{{dx_1}\over{x_1}}
\int{{dx_2}\over{x_2}} f_i(x_1)f_j(x_2)\int dz \int dp_t^2
{{d\sigma}\over{dz dp_t^2}}
\ee
and
\be
\label{qmaxav}
<q_{max}(s)>={{\sqrt{s}} 
\over{2}}{{ \sum_{i,j}\int {{dx_1}\over{ x_1}}
f_{i/a}(x_1)\int {{dx_2}\over{x_2}}f_{j/b}(x_2)\sqrt{x_1x_2} \int dz (1 - z)}
\over{\sum_{i,j}\int {dx_1\over x_1}
f_{i/a}(x_1)\int {{dx_2}\over{x_2}}f_{j/b}(x_2) \int(dz)}}
\ee
with 
    the lower limit of integration in the variable $z$ given by 
 $z_{min}=4p_{tmin}^2/(sx_1x_2)$.
Using for the valence quarks the same approximation
as in the Drell-Yan case, we obtain to leading order
\be
\label{qmaxavan}
<q_{max}(s)>\sim {{3}\over{8}} p_{tmin}
ln{{\sqrt{s}}\over{2p_{tmin}}}
\ee
for $2p_{tmin}<<\sqrt{s}$. For $p_{tmin}=1.4\ GeV$,  
as in typical eikonal mini-jet models for proton-proton scattering
\cite{BLOCK}, one obtains values of  $<q_{max}(s)>$ which range
from 0.5 to 5 GeV for $\sqrt{s}$ between 10 GeV and 14 TeV respectively.
A more precise
 evaluation of the above quantities depends upon
  the type of parton densities one uses, and will be discussed in a
forthcoming paper.
  \begin{figure}[t]
 \begin{center}
\leavevmode
\mbox{\epsfig{file=partonspacefig2.ps,width=0.7\textwidth,bbllx=30pt,bblly=70pt,bburx=570pt,bbury=750pt,angle=90}}
\end{center}

\caption{Comparison between the  b-distribution 
 from the Bloch-Nordsieck model (full) and the  form factor model (dots).}

\end{figure}

\thispagestyle{plain}
 From the discussion about the large b-behaviour of the function $h(b)$,
   we then expect
  $A_{BN}(b,s)$ to fall at large b more rapidly as the energy increases
   from $\sqrt{s}=10 \ GeV$ 
  into the TeV region. 
In Fig. 2, we compare this behaviour for  the function $A(b)$ with the one
obtained through the
  Fourier transform of the squared e.m. form factor of the proton, i.e.
  \be
  \label {AFF}
A(b)=\int {{d^2{\vec Q}\over{(2\pi)^2}}} e^{i {\vec Q}\cdot {\vec b}} 
\big({{\nu^2}\over{Q^2+\nu^2}}\big)^4={{b \nu^2  \sqrt{\nu^2}}
\over{96 \pi}} {\cal K}_{3}(b \sqrt{\nu^2})\ \ \ \ \ \ \ \nu^2=0.71\ GeV^2
\ee
which is the one proposed   by L.Durand et al. \cite{DURAND}
in the first eikonal mini-jet model for proton-proton collisions. The function
$A(b)$ from the Bloch-Nordsieck model is calculated for $\Lambda=0.1\ GeV$
and 
 values of  $<q_{max}>$ which include those obtainable from
 eq.(\ref{qmaxavan}) in the energy  range
$\sqrt{s}\approx 10\ GeV \div 14\ TeV$.

We notice that, as the energy increases, $A(b) $ from 
 the form factor model remains substantially 
higher at large b 
than in the  Bloch-Nordsieck case.
 As a result, for the same 
$\sigma_{jet}$ the Bloch-Nordsieck model
will give smaller $n(b,s)$ at large b 
than the form factor model and
a softening effect of the total 
eikonal mini-jet cross-sections can be expected.
\par\vskip 5 mm

In summary, a soft gluon summation model allows one to obtain 
a  value for the intrinsic transverse momentum cut-off required in
the b-distribution of partons in a proton in agreement with current phenomenological
estimates \cite{PYTHIA}. For such a 
description, an ansatz is necessary for the behaviour of the QCD coupling 
constant in the near zero region. By using suitable
energy scale for the maximum energy allowed for emission of
a single soft gluon by the valence quarks, and borrowing the expression 
for $\alpha_s$ used elsewhere which leads to  calculable non-singular 
integrals, we have obtained 
a progressively softer distribution in the large b-region 
as energy is increased. The 
physical interpretation of this
effect can be traced to the fact that,
as the energy increases,  partons undertake
scattering at smaller and smaller b-distances.
A comparison with the b-distribution from proton form-factor
type models indicates a distinctly different behaviour in this
 large b-region suggesting a  softening of the rise 
 of the total cross-section in mini-jet models relative 
 to the ones with the proton form-factor. Work
  is in progress to apply this model to hadron-hadron
   and photo-hadron collisions.

\thispagestyle{plain}

This Work has been partially supported by CICYT, Contract \# AEN 94-00936,
 EEC HCMP CT92-0026, and US-Department of Energy.
\thispagestyle{plain} 

\end{document}